# Powering Electronic Devices from Salt Gradients in AA Battery-Sized Stacks of Hydrogel-Infused Paper

Anirvan Guha[1], Trevor J. Kalkus[1], Thomas B.H. Schroeder[2], Oliver G. Willis[3], Michael Mayer[1,*]

[1]Adolphe Merkle Institute, University of Fribourg, Fribourg, Switzerland

[2]John A. Paulson School of Engineering and Applied Sciences, Harvard University, Cambridge, MA, USA

[3]Department of Chemistry, Durham University, Durham, UK

[*]Correspondence should be addressed to M.M. (michael.mayer@unifr.ch)

**Strongly electric fish use gradients of ions within their bodies to generate stunning external electrical discharges;[1] the most powerful of these organisms, the Atlantic torpedo ray, can produce pulses of over 1 kW from its electric organs.[2] Despite extensive study of this phenomenon in nature,[1,3] the development of artificial power generation schemes based on ion gradients for portable, wearable or implantable human use has remained out of reach. Previously, inspired by the electric eel, we developed an artificial electric organ that generated electricity from ion gradients within stacked hydrogels and, like the eel, was optimized to deliver large voltages that exceeded 100 V.[4] Due to its high internal resistance, the current of this power source was, however, too low to power standard electronics. Here we introduce an artificial electric organ that takes inspiration from the unique morphologies of torpedo rays for maximal current output. This power source uses a hybrid material of hydrogel-infused paper to create, organize, and reconfigure stacks of thin, arbitrarily large gel films both in series and in parallel. The resulting increase in electrical power by almost two orders of magnitude compared to the original eel-inspired design makes it possible to power electronic devices and establishes that biology's mechanism of generating significant electrical power can now be realized from benign and soft materials in a portable size.**

The ability of strongly electric fish to produce external electrical discharges has inspired the design of artificial power sources since Alessandro Volta's conception of the first battery.[5,6] Researchers have gravitated towards studying members of the *Electrophorus* genus, commonly known as electric eels, due to their exquisite physiological capacity for generating large voltages. For instance, the recently discovered *Electrophorus voltai* can produce 860 V from the aggregate of thousands of electrically active 'electrocyte' cells stacked in series within its electric organ.[1,7] Inspired by the large voltages of electric eels, we recently engineered a soft, flexible, and potentially biocompatible power source from gradients of ions stored within hydrogels.[4] This system mimicked the organization of the eel's electric organ by stacking a repeat sequence of high-salinity, cation-selective, low-salinity, and anion-selective hydrogels in series to produce over 100 V. The maximum power density of 27 mW m$^{-2}$ per repeat unit of this system was, however, too low to power electronic devices. Reaching practically useful power output from an artificial electric organ with the approximate size of a standard AA battery would require a strategy for rapidly fabricating and stacking thin, large-area hydrogel films without curling, tearing, or wrinkling to ensure maximal electrical conduction between adjacent gels.[4]

Here, we took inspiration from the morphological adaptation of torpedo rays to overcome the limitations of the eel-inspired design and to introduce the ability to power small electronics. Specifically, in order to maximize current instead of voltage, we developed a practical method for rapidly stacking extremely thin sheets (< 500 μm) with tunable surface area to assemble electrically active repeating units that are reminiscent of the stacks of electrocytes in the electric organs of torpedo rays (Fig. 1). To this end, we introduce a hybrid material consisting of a thin porous scaffold infused with a hydrogel. The resulting next-generation artificial electric organ offers four enabling advancements over its predecessor:[4]

First, it makes it possible to fabricate, manipulate and stack thin, arbitrarily large hydrogel films in a practical and rapid fashion. Second, it introduces a modular strategy for bringing repeat sequences of these thin gel sheets into near-synchronous contact, thereby enabling reconfigurable serial or parallel arrangements of gel stacks with the ideal number of repeat units and surface area. Third, it provides a straightforward, mechanically supported platform to optimize hydrogel compositions as well as other system parameters. And fourth, it enables planar as well as rolled assembly strategies; rolling is particularly attractive because it is industrially used for paper and thin films, extremely rapid, and automatable.

Both the artificial electric organ and its natural source of inspiration rely on selective movement of ions through aqueous solutions of electrolytes to generate voltage and current (Fig. 1).[2,4] Consequently, a significant portion of the electrical resistance of power generation schemes from ion gradients arises from the resistance of these electrolyte solutions to ion flux, which is directly proportional to the length of the conductive pathway and inversely proportional to its cross-sectional area.[8,9] From this perspective, torpedo rays stand out as examples of electric fish that evolved to generate electric shocks of maximal power. Torpedo electrocytes are thin, polygonal sheets (Fig. 1a), representing the ideal geometry for minimal internal resistance by maximizing cross-sectional area and limiting transport distance. Notably, Volta modeled the thin metal discs of his galvanic battery after the shape of torpedo electrocytes rather than those of electric eels, which are ribbon-shaped.[10] For instance, the Atlantic torpedo *Tetronarce nobiliana* organizes up to 2000 columns of electrocytes in parallel to generate electrical discharges of over 1 kW, a 10-fold higher power output than that of the electric eel.[2,3] Remarkably, the torpedo achieves this elevated power output even though both its individual electrocytes and electric organ as a whole produce significantly lower voltages than those of the eel (Table 1). This performance illustrates the importance of the torpedo's flattened body shape with large area to accommodate a massively parallel arrangement of transverse columns of extremely thin electrically active cells to enable maximum current output.

To mimic the torpedo's 'stack of thin sheets' architecture, we created hierarchical double networks composed of sheets of paper infused with hydrogels of desired composition and function.[11,12] Paper provides the material with mechanical support; we chose it as a scaffold because it is thin, porous, robust, hydrophilic, bendable, foldable, biodegradable, non-toxic, and potentially biocompatible. These qualities contribute to the increasing use of paper in a wide range of applications, including fluidics, robotics, and electrochemical devices and assays.[13–15] To create paper-gel films, we dispensed precursor solutions of each hydrogel composition onto pieces of cellulose paper. The porosity and water-absorbent characteristics of the paper resulted in an autonomous, self-induced wicking action that infused the scaffold rapidly and evenly with hydrogel precursor solutions resulting in thin, coated sheets. Curing by UV-induced polymerization formed gel films attached firmly to the embedded substrate, creating paper-gel hybrid materials similar to those pioneered by the Whitesides group.[16] Such paper-gels improve the mechanical stability compared to unsupported hydrogels, thereby limiting the risk of curling or wrinkling while retaining enough flexibility for bending, rolling or folding during assembly, and these benefits are associated with no apparent losses in ion selectivity or conductivity. To assemble a functioning artificial electric organ, we layered or rolled high-salinity, cation-selective, low-salinity, and anion-selective paper-hydrogel films in the same fashion as one would stack or roll four sheets of paper. Compared to our previous best design of the artificial electric organ,[4] this new design with 500 µm-thick gel films that made close electrical contact over the entire surface of the films improved the power density by one order of magnitude (Fig. 2).

To increase the power further, we performed a rigorous optimization of the choice of charge-selective hydrogel, type of salt, concentration of salt, magnitude of salt gradient and thickness of paper-gels. These studies revealed that sodium chloride, the principal salt of our first artificial electric organ,[4] produced relatively low voltage and power compared to other chloride salts. We identified lithium chloride (LiCl) as the best replacement for three reasons. First, artificial electric organs containing lithium chloride produce a high open-circuit voltage compared to most other salts (Fig. 2b), likely due to differences in ion activity as discussed in Supplementary Information 1. Second, although LiCl did not provide the highest power of all the salts we tested, it is the most soluble in water. Increasing the ionic strength of the solutions decreases the resistance of the system, so the high salt concentrations enabled by using LiCl yield low overall resistances.[17,18] As Figure 2c shows, increasing the concentration of the high-salinity gel using lithium chloride, while maintaining a 100-fold gradient in salt concentration, resulted in a linear increase in power due to a concomitant decrease in resistance. Third, LiCl is hygroscopic, allowing a gel containing 6 M LiCl to retain its entire water content 60 h after curing, while a gel containing 1.5 M NaCl lost more than half of its mass by evaporation during the same period (Supplementary Information 2, Fig. S1).

Using a LiCl gradient in combination with gels supported by a scaffold of lens-cleaning paper that was four times thinner than the chromatography paper we initially employed, resulted in a peak power output of 1.80 W m$^{-2}$ per tetrameric gel cell. This performance represents a 67-fold improvement over our previously published best configuration (Fig. 2e, f)[4] and is only a factor of 5 to 15 smaller than the power density of the strongest electric fish such as the electric eel or torpedo ray (Table 1). Other power sources that employ ionic gradients to generate similar power densities either range in size from benchtop devices to technical scale pilot plants and are therefore too big to conceive possible future implantation,[19,20] or – if they are sufficiently small – they can at present only deliver nanoampere currents because constraints in membrane fabrication and robustness make scaling to large cross-sectional areas currently impossible.[21–23] In contrast, the paper-gel electric organ provides a finger-sized (Fig. 3 and Fig. S3), potentially biocompatible and mechanically soft platform that can generate one million-fold larger currents in the range of milliamperes (Fig. 2e) from benign components that include water, salt, hydrogel, and paper.

Figure 3 demonstrates that these improvements in power density make it possible to reach an essential milestone towards portable, wearable, and implantable electrical power sources; they render artificial electric organs driven solely by ion gradients capable of powering real-world electronic devices. Moreover, they provide this power in a format of only $1.5 \times 1.5 \times 3.2$ cm, corresponding to a volume of less than 8 mL, which is sufficiently small to envision future implantation. We used such a stack of 16 repeating units in series to power a circuit with a pre-programmed microcontroller chip that illuminated several LEDs in sequence (Fig. S2). This result, together with Supplementary Video 1 and Supplementary Information 3, unequivocally establish that a relatively small stack of scaffolded hydrogels can power sequences of logical operations within a computer. Figure 3 also illustrates that straightforward reconfiguration of the dimensions of a paper-gel stack makes it possible to meet a range of operational demands for various electronic devices with low power consumption. For instance, while eight gel cells connected in series (volume = 3.6 mL) were able to illuminate a red LED, doubling the number of cells in series or in parallel doubled the open circuit voltage or the cross-sectional area of the stack and made it possible to illuminate three LEDs of different colors, powered solely by the paper-gel system. Modularity like this is beneficial for cases where either a threshold current or, as in the case of a green LED, a threshold voltage must be reached to operate various devices.

This modularity was made possible by a straightforward assembly and reconfiguration strategy as illustrated in Figure 4. Specifically, we took advantage of rapid and controlled tearing, bending, and folding along laser-cut perforations to create and stack a desired number of layers of large-area tetrameric gel cells (Fig. 4a). Figure 4b demonstrates that, even after infusing the paper scaffold with hydrogel and curing, the laser-cut subdivisions facilitate rapid tearing of tetrameric gel layers followed by restacking individual paper-gel cells to many cells in series or in parallel. In order to explore if paper-gel systems could potentially be assembled using a well-established industrial process, we took advantage of the mechanical flexibility of the paper-gel layers and rolled a long strip of the four different paper-gels layered on top of each other (Fig 4c). This process rapidly created repeating tetrameric sequences with each revolution. Figure 4c shows that cutting one such roll perpendicular to the direction of the rolling motion removed the lateral connection between the sheets and created a stack of three large-area gel cells in series. Industrial rolling machines can wind over one kilometer of paper per minute;[24] applying this process to four-layered paper-gels with a width of 2 m could, within seconds, create a stack of 360 gel cells to generate at least 1 kW power. While the power of such stacks would dissipate rapidly upon contact between the gel sheets, we have demonstrated previously that applying a current through a discharged assembly of gels recharges these devices to recover and even exceed their maximum power.[4]

The innovations required to transform the artificial electric organ from a high-impedance scheme[4] to a suitable power source for electronic devices recall the distinct evolutionary paths that resulted in electric eels and torpedoes. Achieving maximum power transfer requires matching the internal impedance of a power source with its load impedance.[25] Since the 19th century, researchers have postulated that this principle explains the evolution towards the long and relatively narrow shape of the electric organs found in electric eels such that their internal resistance matches the low-conductivity environment of fresh water. Conversely, the strong electric organs found in fish native to high-conductivity salt water evolved in flat, wide rays that provided the development of organs with a low internal resistance.[26] Contemporary portable electronic devices frequently have operating voltages of less than ten volts and are powered by batteries with sub-ohm output impedances.[27] As powering such devices more closely approximates the evolutionary imperative of the torpedo than that of the eel, we found that reaching the power output necessary for doing useful electrical work required redesigning the artificial electric organ from the eel-inspired, long arrangement of electrically active cells in series for high voltage generation to the torpedo-inspired, flat, wide, and parallel arrangement for high current generation.

One remaining barrier to this system's utility as a practical power source is, however, its inability to maintain maximal power output for extended periods of time. Once a tetrameric sequence of gels comes into contact, the salt gradient responsible for generating the potential begins to dissipate. The rate of this dissipation increases when the organ is connected to a load to do electrical work. For instance, a LED powered by a stack of paper-gel cells lost 90% of its maximum brightness within 5 min (Supplementary Information 4, Fig. S4). Power-on-demand devices for single-use electronics, such as paper-based diagnostics,[15,28] may, however, constitute applications that are not affected by this limitation. These kinds of disposable point-of-care devices are often required in low-resource areas where traditional batteries are prohibitively expensive and their disposal may endanger the environment.[29] The amount of power required for these devices, as well as for certain electrochemical or fluorescent diagnostic assays, is well within the reach of the paper-gel electric organ.[30,31]

For the successful application of paper-gel cells in such low-cost, non-toxic, single-use electrical devices, it would be critical that they can be dehydrated, stored without dissipation of the gradient, and activated on demand. To explore this possibility, we dehydrated individual paper-gel films, stored them for one day and then rehydrated the films before stacking them to an electrically active gel cell. While the initial output power of this rehydrated artificial electric organ was only about half that of a freshly prepared one, its power output remained almost constant over 5 min. By this time, the ionic gradient of the freshly prepared artificial organ had partially dissipated, resulting in approximately half of its initial power and hence both versions provided comparable power output (Supplementary Information 5, Fig. S5). Considering that all components of these paper-gel electric organs are environmentally friendly and potentially biodegradable,[32,33] these results hint at the artificial organ's promise as a safely disposable power source that delivers electricity on demand by a process as simple as adding water or other readily available aqueous solutions.

Strongly electric fish show that the energy stored within ionic gradients, can generate powerful and useful discharges of electricity within the constraints of living organisms. The work presented here establishes a straightforward and robust method of powering electronic devices from rapidly assembled stacks of ionic gradients. Since the total volume of these potentially biocompatible hydrogel stacks is smaller than 10 mL and would readily fit into a body cavity, and since the peak power density of this paper-gel electric organ's represents roughly $1/10^{th}$ of the power density of the most powerful electric fish, we suggest that the vision of implantable artificial electric organs has made one more encouraging step towards realization. Moreover, with further improvements, artificial electric organs have the potential to reach and possibly exceed the performance of biological electric organs. These improvements may include the use of thinner scaffolds and gel films, further optimized hydrogel compositions and polymerization conditions, or laminating the high and low-salt compartment gels with ultra-thin films,[34,35] molecular layers,[36,37] or 2-D materials[21] that are capable of charge- or even ion-selective permeation. Even without these improvements, however, the work presented here establishes that artificial electric organs from stacks of paper-gels can do useful electric work, a feat that was out of reach of our first artificial electric organ presented recently.[4] If future designs can maintain or regenerate ion gradients, perhaps by exploiting gradients that are physiologically maintained in humans such as the proton concentration in the stomach versus fluids in the surrounding tissue or by generating ions in situ from products of human metabolism,[38] then the application of the paper-gel electric organ may expand to powering portable, wearable, and possibly even implantable electronics.

**Acknowledgements**

We thank Jerry Yang for discussions, critical comments on the manuscript and suggestions. We are grateful to U. Steiner's group for the use of their Formlabs 3D Printer. Laser cutting was performed at Fablab Fribourg. Research reported in this publication was supported by the National Center for Competence in Research (NCCR) from the Swiss National Science Foundation (SNSF) on Bioinspired Materials and by a PIRE grant on Bio-Inspired Materials and Systems funded jointly by the National Science Foundation of the US (NSF) and the SNSF. T.B.H.S. is funded by a SNSF Postdoc.Mobility fellowship. We are also thankful to the Adolphe Merkle Foundation for support.

**Author Contributions**

A.G. and M.M. conceived the project. A.G., T.J.K. and M.M. designed the experiments. A.G., T.J.K. and O.G.W. collected all data. T.B.H.S. provided analysis on the impact of salt choice on the power output of the artificial electric organ and linked the innovations presented here to the divergent evolutionary paths of electric eels and torpedoes. A.G., T.B.H.S. and M.M. wrote the manuscript.

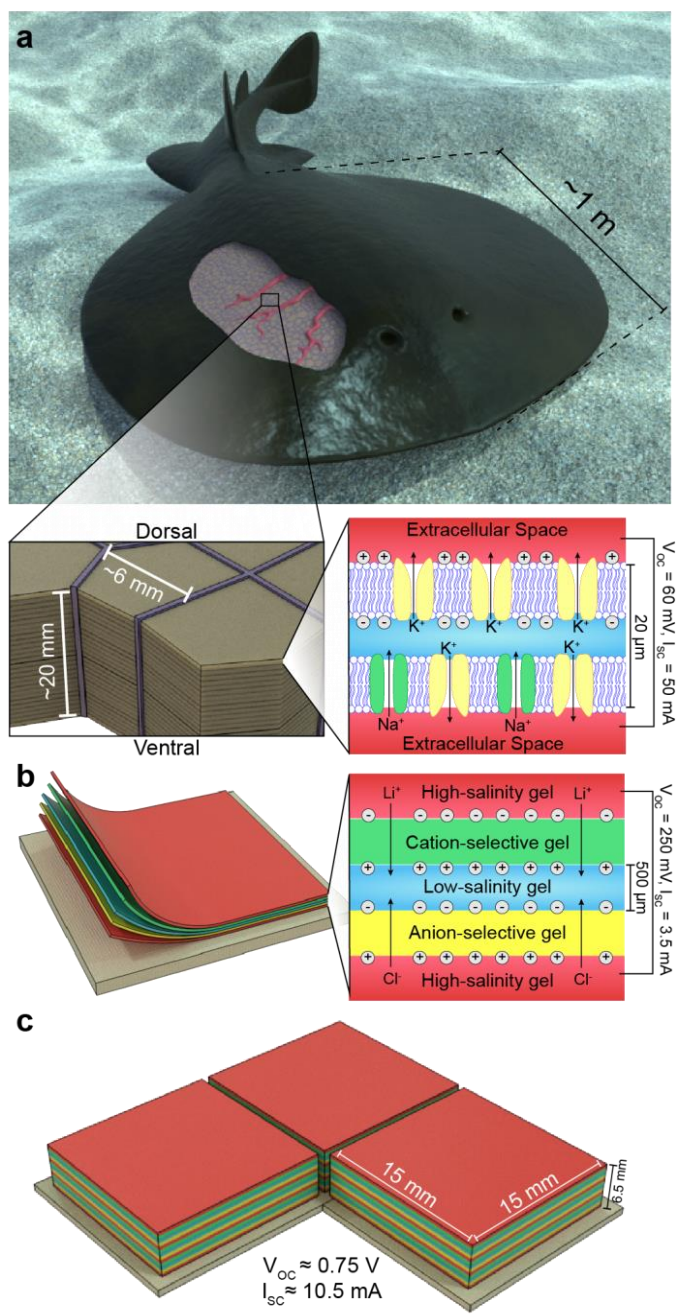

**Figure 1. Anatomy of the electric organ of the ray *Tetronarce nobiliana* and design of a paper-gel electric organ. a,** *Tetronarce nobiliana* has a flat, disc-shaped body up to one meter in diameter. A pair of electric organs are located on either side of the ray's body. Each organ is a parallel arrangement of transverse columns consisting of hundreds to thousands of electrocytes in series (inset, left). The organization of these thin, large-area cells is reminiscent of stacks of thin sheets. The electrocytes here are drawn out of scale for clarity; the small

thickness of the cells and small spacing between adjacent cells would make them impossible to distinguish from each other if they were drawn to scale. When activated by the ray's nervous system, each electrocyte generates a transcellular potential of ~60 mV from the asymmetrical flux of sodium and potassium through ion-selective channel proteins in its cell membranes (inset, right). **b,** Principle of the paper-gel electric organ. Sheets of paper, infused with hydrogel, are stacked in a repeating tetrameric sequence to generate additive potentials (inset). The red hydrogel is polymerized from uncharged monomers and stores a concentrated salt solution. The green hydrogel contains negatively charged monomers and is selective for permeation of cations. The blue hydrogel contains uncharged monomers and dilute salt solution. The yellow hydrogel contains positively charged monomers and is selective for permeation of anions. **c,** Stacking these tetrameric gel cells in series and in parallel linearly scales voltage and current, respectively.

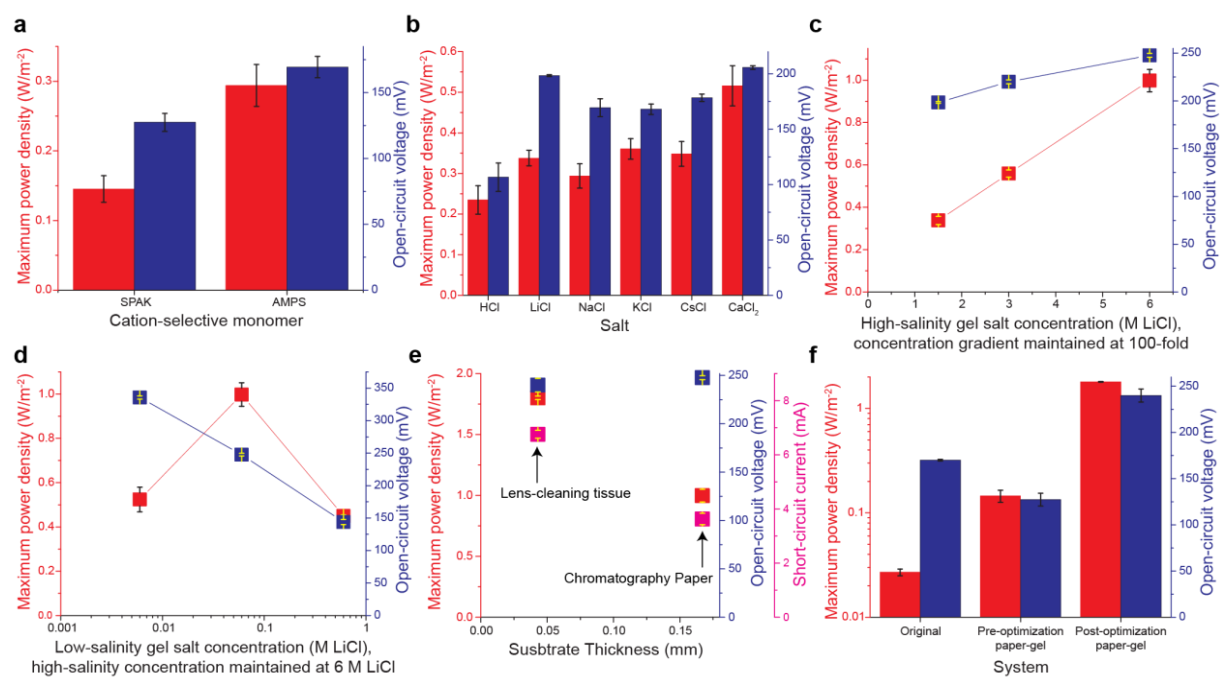

**Figure 2. Optimization of the paper-gel electric organ for maximal voltage and power output.** Each test was performed by stacking one sequence of high-salinity, cation-selective, low-salinity, anion-selective, and high-salinity paper-gels and measuring the resulting maximum power density (red) and open-circuit voltage (blue). All data are presented as mean ± s.e.m.; $n = 3$ for all experiments except for lens-cleaning tissue data from panel e, where $n = 2$. **a,** Comparison of two cation-selective monomers for hydrogel formation. Using 2-acrylamido-2-methylpropane sulfonic acid (AMPS) instead of 3-sulfopropyl acrylate (potassium salt) (SPAK) increased open-circuit voltage and doubled maximum power output. **b,** Influence of the type of chloride salt in the high- and low-salinity gels on electrical output. **c,** Increasing the salt concentration in the high and low-salinity gel linearly increased both voltage and power output. **d,** Increasing the concentration gradient between the low- and high-salinity gels increased the voltage, but decreasing the low-salinity gel concentration to achieve a larger gradient resulted in a decrease in overall power output. **e,** Using 4-fold thinner lens-cleaning tissue as the paper substrate rather than chromatography paper nearly doubled the short-circuit current (pink) and power output of the paper-gel stack. **f,** Comparison between the original artificial electric organ, the paper-gel electric organ with similar hydrogel compositions as the original artificial electric organ, and the paper-gel electric organ after optimizing hydrogel compositions, type and concentration of salt, and substrate thickness to maximize power output.

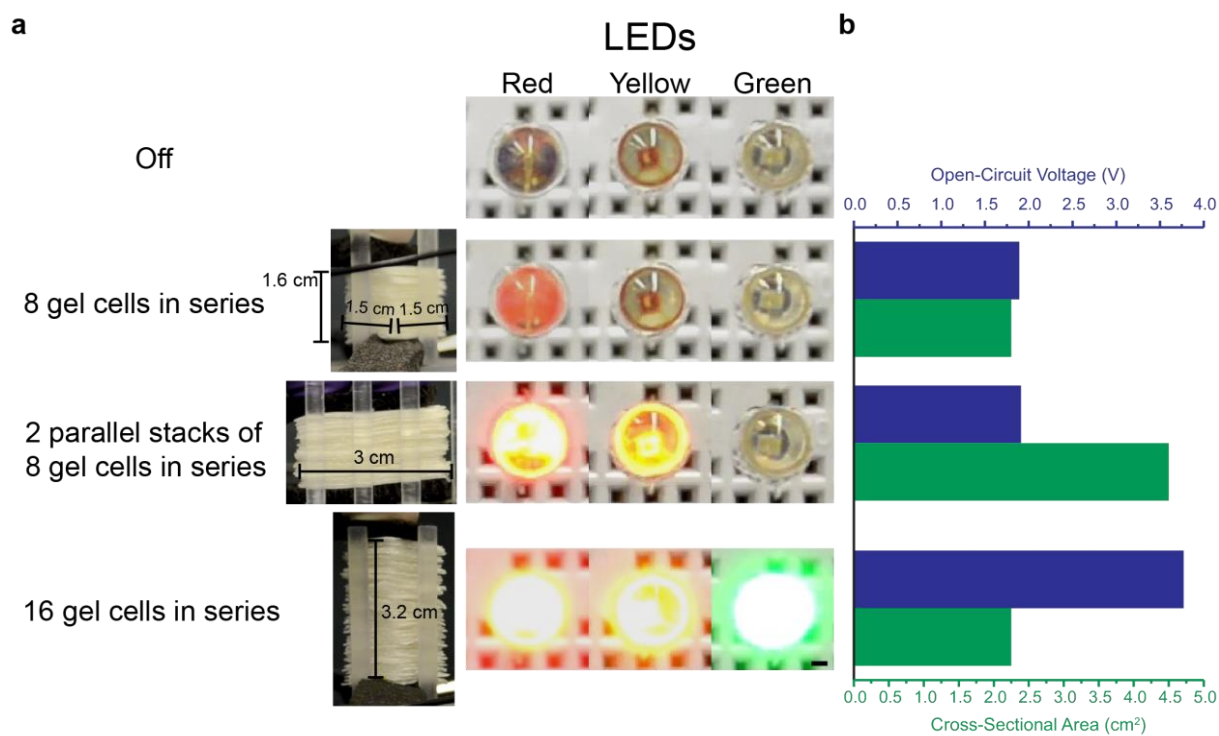

**Figure 3. Powering LEDs with paper-gel cells. a,** Photographs of red, yellow, and green LEDs depict their luminosity when connected in series with different stacks of paper-gels. When eight paper-gel cells in series were used as the power source, only the red LED was dimly lit. Adding a second stack with eight paper-gel cells in parallel or extending a single stack from 8 to 16 paper-gel cells in series increased the overall power output of the stacks to 1.6 mW and made it possible to power all the LEDs with increased brightness. The yellow LED required a minimum current of 32 μA to turn on, which was reached with two stacks of 8 gel cells connected in parallel. The green LED had a forward voltage of 3.2 V, so it only illuminated after increasing the open-circuit voltage of the stack to near or above 3.2 V. A photograph of the complete circuit is shown in Figure S3. Scale bar, 1 mm. **b,** Open-circuit voltage (blue bars) and cross-sectional area (green bars) of the adjacent stacks from panel a.

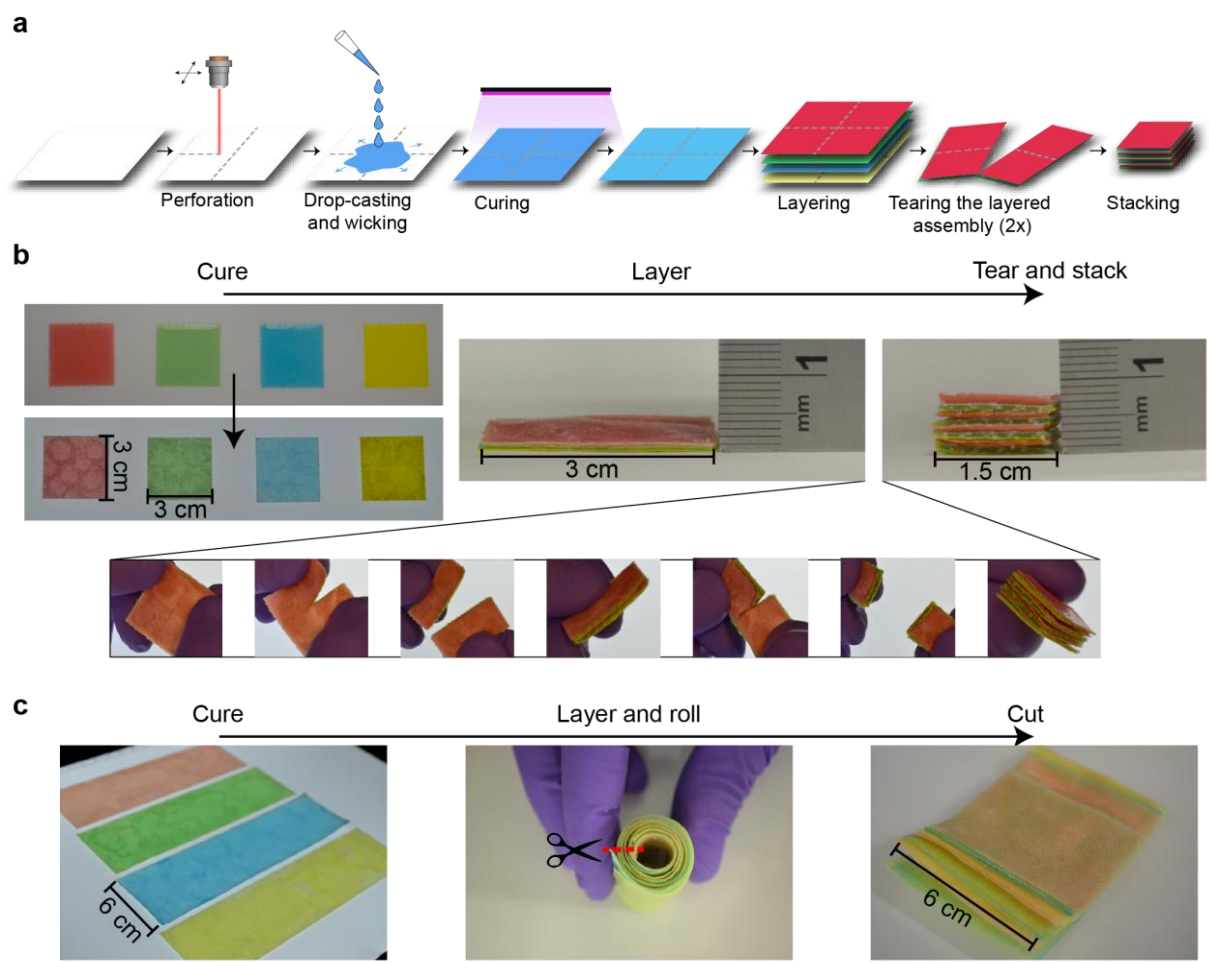

**Figure 4. Paper-gel electric organ supports multiple architectures for rapid assembly of stacked hydrogel films.** Dye was used for clarity to visualize the four different hydrogel compositions (red = high-salinity gel; green = cation-selective gel; blue = low-salinity gel; yellow = anion-selective gel). **a,** Illustration of the process of converting sheets of cellulose paper to an electrically active stack of paper-gels. The paper was perforated using a commercial laser cutter. Hydrogel precursor was drop-cast onto each perforated sheet of paper and the wicking action of the paper spontaneously coated it in a thin film of liquid, which was cured with a UV light. Sheets of the four hydrogel compositions were layered into a stack, which was then torn along the overlaid perforations. The resulting subunits were restacked to form a stack of several tetrameric paper-gel cells in series. **b,** A straightforward tearing and stacking method converted one paper-gel cell into four cells in series following the procedure depicted in panel a. The inset shows how a single stack was torn along its perforations and restacked twice to

quadruple the number of cells stacked in series. **c,** Rolling and then cutting one paper-gel cell produced three large-area repeat units in series: After creating the four different large-area paper-gels (left) they were layered in sequence and then rolled three times around a rod (middle). The roll was cut along the indicated cut line and unfurled to produce three large-area paper-gel cells in series (right).

**Table 1: Comparison of geometrical and electrical characteristics of two types of natural and two types of artificial electric organs**

| Source | Thickness of cell* (m) | Cross-sectional area of cell (m$^2$) | Cross-sectional area of electric organ† (m$^2$) | Ratio of cell cross-sectional area to cell thickness (m) | Ratio of organ cross-sectional area to cell thickness (m) | Open-circuit voltage per cell (V) | Internal resistance per cell (Ω m$^2$) | Maximum power density generated by cell (W m$^{-2}$) |
|---|---|---|---|---|---|---|---|---|
| *Electrophorus electricus*[1,4,39–41] | $(1.0 \pm 0.2) \times 10^{-4}$ | $6 \times 10^{-5}$ | $(2.8 \pm 0.6) \times 10^{-3}$ | 0.60 | 28 | $0.13 \pm 0.01$ | $(5.6 \pm 0.6) \times 10^{-4}$ | $9.9 \pm 1.7$ |
| *Tetronarce nobiliana*, Bennett *et al.* 1961§ | $2.0 \times 10^{-5}$ | $2.8 \times 10^{-5}$ | $4.2 \times 10^{-2}$ | 1.4 | 2100 | 0.06 | $3.2 \times 10^{-5}$ | 28.3 |
| Gel cells, 80° Miura-ori fold, Schroeder *et al.* 2017 | $2.8 \times 10^{-3}$ | $8.5 \times 10^{-5}$ | | 0.03 | | $0.17 \pm 0.01$ | $0.27 \pm 0.02$ | $0.027 \pm 0.002$ |
| Optimized paper-gel cells, this work | $1.4 \times 10^{-3}$ | $2.25 \times 10^{-4}$ | | 0.16 | | $0.24 \pm 0.01$ | $(8 \pm 1) \times 10^{-3}$ | $1.8 \pm 0.01$ |

Where applicable, all values are presented as mean ± s.e.m.

\* Cell refers to an electrocyte in the natural electric organs and a tetrameric gel cell in the artificial electric organs.

† For *T. nobiliana*, the areas of both electric organs were added together.

‡ Values presented here are the averages of the three studies[39–41] on *Electrophorus electricus* presented in Table 1 of the electric-eel-inspired artificial electric organ manuscript.[4] Only the cross-sectional area of an *Electrophorus* electrocyte is added here. From Gotter,[1] the average electrocyte was assumed to be a 40 mm wide and 1.5 mm tall rectangular prism.

§ Parameters are given for a 1 m diameter specimen of *T. nobiliana*. Bennett provides ranges for several parameters; in such cases, the average of the extremes was used for calculations. Assumptions: *T. nobiliana* contains 1500 parallel columns of 1000 electrocytes in series. The shape of each electrocyte was approximated as a disc, with an average diameter of 6 mm and an average thickness of 20 µm.

**Methods**

**Materials and equipment**

We purchased all chemicals from Sigma-Aldrich (Merck-KGaA) except for 40% (w/v) 37.5:1 acrylamide/N,N′-methylenebisacrylamide (henceforth 'bis') solution (Bio-Rad). We purified water to 18.2 MΩ cm with a PURELAB Flex II purifier (ELGA LabWater, Veolia). We used either grade 1 chromatography paper (thickness = 0.17 mm) (Whatman plc) or MC-5 lens-cleaning tissue (thickness = 0.04 mm) (Thorlabs Inc.) as paper substrates. We perforated these paper substrates using a Speedy 300 laser cutter (Trotec) or a Cricut Maker cutting machine (Cricut Inc.) equipped with a perforation blade. We cured all gels with a Mineralight UV Display lamp (UVP, Analytik Jena) containing two 25-W, 302-nm UV tubes (Ushio Inc.). We used graphite felt (SGL Carbon GmbH) or chlorinated silver foil (Sigma-Aldrich) as electrodes. We purchased all electronic components through Distrelec Group AG. We 3D-printed supports for stacks of paper-gels (such as the one pictured in Figure 3) with a Form 2 3D printer (Formlabs).

**Composition of hydrogel precursor solutions**

All hydrogel precursor solutions were aqueous, and made with the following compositions. High-salinity gel: at least 1.5 M salt, 4.38 M acrylamide, 0.054 M bis, and 0.005 M 2-hydroxy-4′-(2-hydroxyethoxy)-2-methylpropiophenone (henceforth 'photoinitiator'). Low-salinity gel: at least 0.015 M salt, 4.38 M acrylamide, 0.054 M bis, and 0.005 M photoinitiator. Cation-selective gel: 2 M of either 3-sulfopropyl acrylate (potassium salt) (SPAK) or 2-acrylamido-2-methylpropane sulfonic acid (AMPS), 3.29 M acrylamide, 0.040 M bis, and 0.005 M photoinitiator. Anion-selective gel: 2 M (3-acrylamidopropyl) trimethylammonium chloride, 2.74 M acrylamide, 0.034 M bis, and 0.005 M photoinitiator. We added food dye from Städter only for photography and left it absent during experiments intended for electrical characterization.

**Fabrication of paper-gel cells**

To create paper-gel hybrids, we placed pieces of paper onto a Teflon sheet. We pipetted 45 µL of the desired hydrogel precursor solution per 1 cm$^2$ of paper, allowing it to wick through the paper until a thin film evenly coated the paper's surface. For papers with an area of 9 cm$^2$ or larger we took care to evenly distribute the liquid across the surface of the paper to expedite the wicking process. We cured the gels with a UV light at a distance of 12 mm for 45 s. We then peeled the paper-gels away from the Teflon and stacked them in a repeating sequence of high-salinity gel, cation-selective gel, low-salinity gel, and anion-selective gel. The first and last gel of every paper-gel sequence was a high-salinity gel.

**Electrical characterization of paper-gel electric organ**

We prepared graphite felt electrodes by dispensing 67 µL of an aqueous solution of 100 mM potassium hexacyanoferrate(II), 100 mM potassium hexacyanoferrate(III), and 1 M KCl per 1 cm$^2$ of felt in the area that would come into contact with the terminal gels; this solution participated in the oxidation-reduction reaction that converted the charges carried by ions within the gels to electrons within the rest of the circuit.[42] We placed these prepared graphite felt electrodes in contact with either end of a stack of paper-gels and applied a small, consistent amount of pressure to ensure even and continuous contact. We recorded voltages with a Tektronix DMM4040 digital multimeter set to high input impedance mode. To calculate maximum power density, we measured the voltage across a fixed-value resistor connected in series with a gel stack (see Supplementary Information 6).

**Optimization of paper-gel system**

For each experiment we prepared an individual paper-gel cell; that is, one sequence of high-salinity, cation-selective, low-salinity, anion-selective, and high-salinity paper-gels. The substrate of each gel was a 1.5 × 1.5 cm square of paper. We used chromatography paper for all tests except where noted. We used a wooden clothespin to hold the electrodes in contact with the paper-gel stack, and measured the open-circuit voltage and maximum power density of each stack. The composition of the gels used in each experiment were as follows, organized by panel from Figure 2:

a. High-salinity gel: 1.5 M NaCl

   Cation-selective gel: 2 M SPAK or 2 M AMPS

   Low-salinity gel: 0.015 M NaCl

   Anion-selective gel: 2 M APTAC

b. High-salinity gel: 1.5 M HCl, 1.5 M LiCl, 1.5 M NaCl, 1.5 M KCl, 1.5 M CsCl, or 1.5 M $CaCl_2$

   Cation-selective gel: 2 M AMPS

   Low-salinity gel: 0.015 M HCl, 0.015 M LiCl, 0.015 M NaCl, 0.015 M KCl, 0.015 M CsCl, or 0.015 M $CaCl_2$

   Anion-selective gel: 2 M APTAC

c. High-salinity gel: 1.5 M LiCl, 3 M LiCl, or 6 M LiCl

   Cation-selective gel: 2 M AMPS

   Low-salinity gel: 0.015 M LiCl, 0.03 M LiCl, or 0.06 M LiCl

   Anion-selective gel: 2 M APTAC

d. High-salinity gel: 6 M LiCl

   Cation-selective gel: 2 M AMPS

Low-salinity gel: 0.006 M LiCl, 0.06 M LiCl, or 0.6 M LiCl

Anion-selective gel: 2 M APTAC

e.  High-salinity gel: 6 M LiCl

Cation-selective gel: 2 M AMPS

Low-salinity gel: 0.06 M LiCl

Anion-selective gel: 2 M APTAC

**Tearing and restacking method of assembling paper-gel cells in series**

To stack many paper-gel cells rapidly in series, we dispensed hydrogel precursor solution onto large pieces of chromatography paper, with perforations for the requisite number of $1.5 \times 1.5$ cm squares. For example, if we wanted to stack four paper-gel cells with dimensions of $1.5 \times 1.5$ cm per paper-gel, we would start with one $3 \times 3$ cm piece of paper for each gel type, with perforations dividing the papers into four $1.5 \times 1.5$ cm squares. After curing, we stacked these large-area paper-gels into one sequence of high-salinity gel, cation-selective gel, low-salinity gel, and anion-selective gel before tearing along the perforations to create two stacks of equal size. We would then stack these two halves in series and tear again, repeating the process to achieve rapidly the desired number of cells in series.

**Powering LEDs with paper-gel stacks**

To power LEDs, we used the tearing and restacking method to create at least eight paper-gel cells in series. The substrate of each paper-gel was a $1.5 \times 1.5$ cm square of chromatography paper. We measured the resulting open-circuit voltage as described previously, and then connected the stack in series with red (Sloan L5-R91H), yellow (RND 135-00030), and green (RND 135-00177) LEDs, one at a time. We used a force-sensitive resistor (FS402, Interlink) to monitor and apply a constant pressure of approximately 6.5 kPa to each paper-gel stack during all measurements and pictures.

**Creating large-area repeating tetrameric paper-gel sequences by rolling**

To provide a rapid and potentially automatable procedure for the fabrication of paper-gel cells with large contact area, we cut a rectangle of chromatography paper with a width of 6 cm for each hydrogel type. The lengths of each rectangle were as follows: high-salinity gel, 13.8 cm; cation-selective gel, 14.5 cm; low-salinity gel, 15.1 cm; anion-selective gel, 15.7 cm. After fabricating each paper-gel as previously described, we stacked the rectangles such that one of the short edges of each rectangle was in perfect alignment, and the opposite end of the rectangles were offset from each other to account for the difference in circumferential path length for the different gels. We pressed the offset end onto the surface of a 0.8 cm diameter rod and rolled the gels around the rod. Once the gels had been rolled fully, we pulled them off the rod and made a cut along the leading edge of the layered paper-gels down to the center of the roll to create three independent tetrameric gel cells.

**Supplementary Information 1.** Effects of electrolyte identity on power generation

As shown in Figure 2b, we found that changing the identity of the type of chloride salts (HCl, LiCl, NaCl, KCl, CsCl, $CaCl_2$) used to create a gradient between high- and low-salinity paper-gels had a significant impact on both the open-circuit voltage and the maximum power density of paper-gel stacks. We propose that these different outputs generated by different electrolytes can be explained by a combination of effects.

Hydrochloric acid yielded the lowest open-circuit voltage and maximum power density among the salts. This can be explained, as the sulfonate group in poly(AMPS) has a p$K_a$ of around 1.5.[1] When AMPS-based membranes are in contact with a 1.5 M solution of a strong acid (i.e. a solution of negative pH), it is reasonable to expect a large proportion of the sulfonate groups in the polymer to be protonated and uncharged, thereby greatly reducing the selectivity of the cation-selective membranes in the system.

Calcium chloride (CaCl$_2$) yielded the highest open-circuit voltage and maximum power density of all the salts. This, too, can be explained: as a 1.5 M solution of CaCl$_2$ has a substantially higher ionic strength than a 1.5 M solution of a monovalent salt, it follows that the electrochemical driving force down a concentration gradient is higher.

The alkali metal chloride salts tested produced fairly similar values of $V_{OC}$ and $P_{max}$, although LiCl had a notably higher $V_{OC}$ than the rest, while NaCl's $P_{max}$ was notably lower than the others. Variation between these salts can be explained by a combination of nonideal solution behavior in the paper-gels containing concentrated electrolyte and differences in ion mobility between salts.[2] The activity coefficients of alkali metal chlorides in aqueous solution decrease when moving down on the periodic table from Li toward Cs, with the largest difference in activity coefficients appearing between lithium and sodium.[3] This trend provides a reasonable explanation for the relatively large electrical potential associated with lithium chloride gradients. One might expect $P_{max}$ to follow a similar trend; however, the high aqueous ionic mobilities of Cs$^+$ and K$^+$ relative to Na$^+$ and Li$^+$ likely enhance their relative maximum power values, bringing their $P_{max}$ values closer to that of LiCl.[4]

**Supplementary Information 2. Dependence of water retention of gels on salt content**

Hydrogels are, by their nature, susceptible to water loss over time due to evaporation. As the water content of a hydrogel of low salinity decreases, so too does its conductivity due to the decrease in mobility of the charge-carrying ions contained within the gel network.[5] Although it may be possible that dehydrating a gel of high salinity could cause it to pass through a regime where its conductivity increases due to the increase in concentration of the salt contained within, we have observed that this effect, if present, does not prevent a steep drop in the conductivity of an entirely dried hydrogel. As its conductivity decreases, the electrical resistance of a hydrogel increases. From the maximum power transfer theorem,[6] the maximum power output $P_{max}$ (W) of a hydrogel power source will be inversely proportional to its internal resistance $R_{int}$ (Ω),

$$P_{max} = \frac{V_{OC}^2}{4R_{int}} \tag{S1}$$

where $V_{OC}$ (V) is the open-circuit voltage of the system. Therefore, for a paper-gel artificial electric organ to generate maximum power for hours or days after paper-gel fabrication, it would need to use hydrogels that are able to retain their water content.

To investigate the water retention of several high-salinity gels, we pipetted 1 mL of 1.5 M NaCl, 1.5 M LiCl, and 6 M LiCl hydrogel precursor solutions onto the surface of 3 separate petri dishes. We cured these hydrogels under a UV light for two minutes, and then recorded the mass of each. After leaving the gels to rest in a fume hood for 60 h, we recorded each gel's mass once again. Although both 1.5 M NaCl and 1.5 M LiCl gels lost over 50% of their mass within this timeframe, the 6 M LiCl gel retained its entire water content (Fig. S1). This result is explained by two, likely interconnected factors: 1) LiCl is known to be a highly hygroscopic salt,[7,8] and 2) increased solubility of salt within an aqueous solution corresponds to a decreased vapor pressure of that solution.[9] This result suggests that a high concentration of LiCl within a gel may allow the decoupling of the gel fabrication process from its use within a power generation system. Although this strategy may only be suitable for the system's high-salinity gels, including a non-ionic humectant (such as glycerol) could increase the water retention of other gels without sacrificing the ionic gradient responsible for voltage generation.[10]

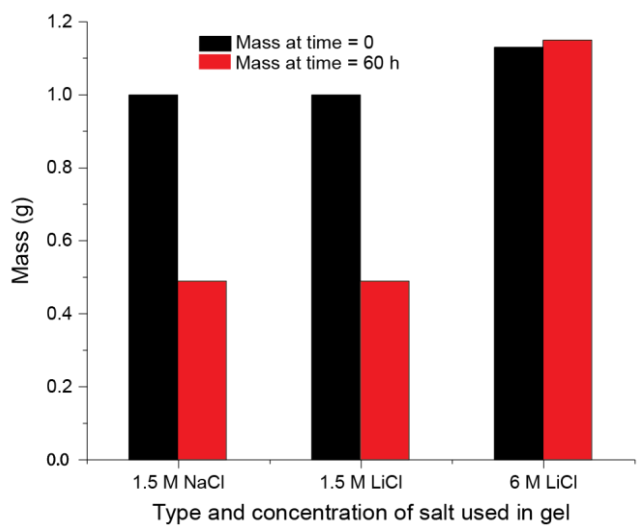

**Figure S1. Water retention of hydrogels over time.** The mass of three different gels is shown shortly after curing (black) and after 60 hours (red).

**Supplementary Information 3. Powering a pre-programmed microcontroller using a paper-gel stack**

To illustrate the potential applications of the paper-gel artificial electric organ as a power source, we programmed an ATTiny84 microcontroller (Microchip Technology Inc.) to light up a series of three red LEDs in sequence repeatedly when connected to a power source. We wrote the code using the Arduino IDE (Arduino Software). After uploading the code to the microcontroller, we detached it from all external power sources. We then assembled a paper-gel stack of 16 cells in series (identical to the bottom stack of Figure 3) and used it to power the microcontroller (Fig. S2). As demonstrated in Supplementary Video 1, we could power both the microcontroller and the LEDs connected to it with the paper-gel stack as the only power source.

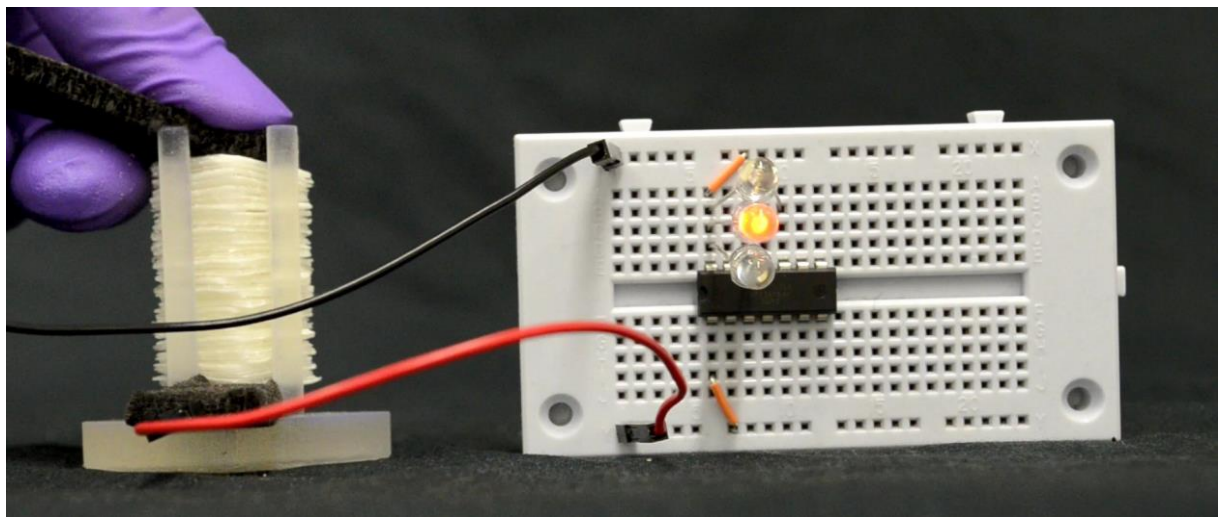

**Figure S2. Microcontroller circuit powered by paper-gel stack.** The photograph shows a circuit containing an ATtiny84 microcontroller connected to three red LEDs and powered by a stack of 16 paper-gel cells in series. The microcontroller was programmed to sequentially redirect current towards three different analogue input pins. Each of these pins was connected to one of three red LEDs. The paper-gel stack delivered enough power to simultaneously power the microcontroller and light up the LEDs in sequence (Supplementary Video 1).

**Supplementary Information 4. Decay of brightness of green LED powered by paper-gel stack**

To evaluate the longevity of the paper-gel artificial electric organ in terms of its use as a power source for electronic devices, we connected a stack of 16 paper-gel cells in series (identical to the bottom stack of Fig. 3) to a green LED (Fig. S3). We directed the LED towards the sensor of a lux meter (Luxmeter LX10, Voltcraft) and measured the illuminance over five minutes (Fig. S4). We chose to illuminate a green LED because its high brightness was easy for the lux meter to detect.

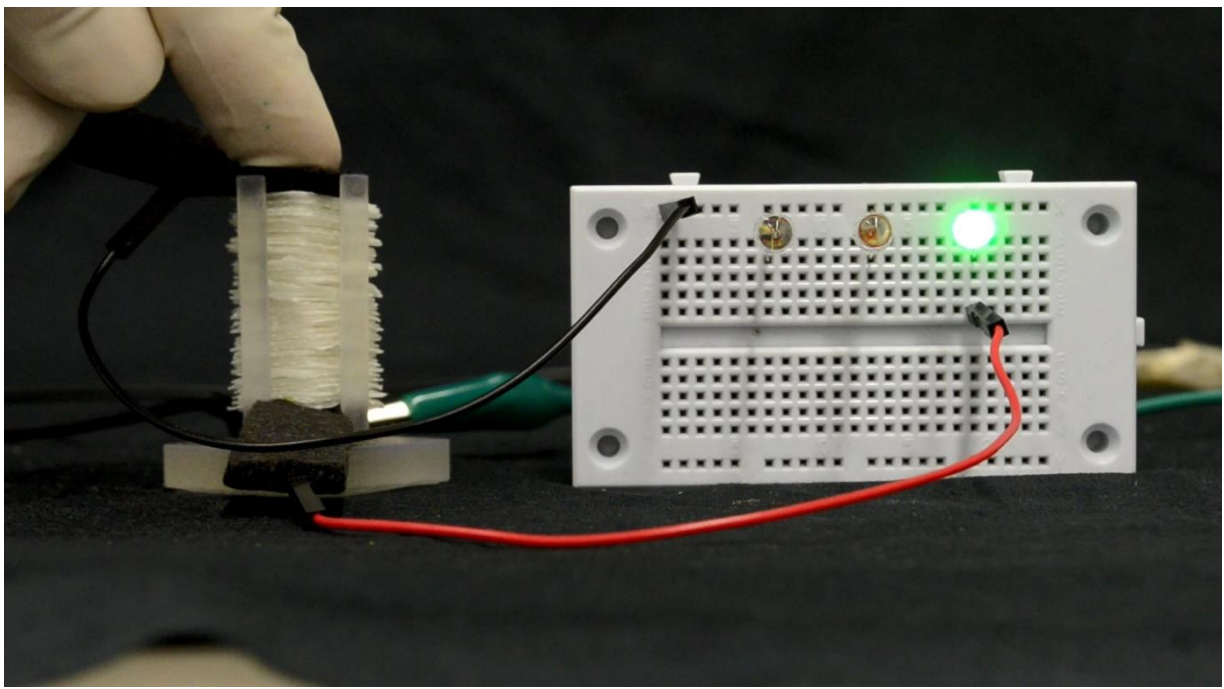

**Figure S3. Full circuit of paper-gel stack powering individual LEDs.** The photograph depicts 16 paper-gel cells (cross-sectional area = 2.25 cm$^2$) in series powering a green LED. A force-sensitive resistor was placed at the bottom of the stack to monitor the pressure applied to the stack. All LEDs were connected to the negative end of the paper-gel stack via the black wire. The red wire was used to connect the positive end of the stack to the positive terminal of each LED, one at a time.

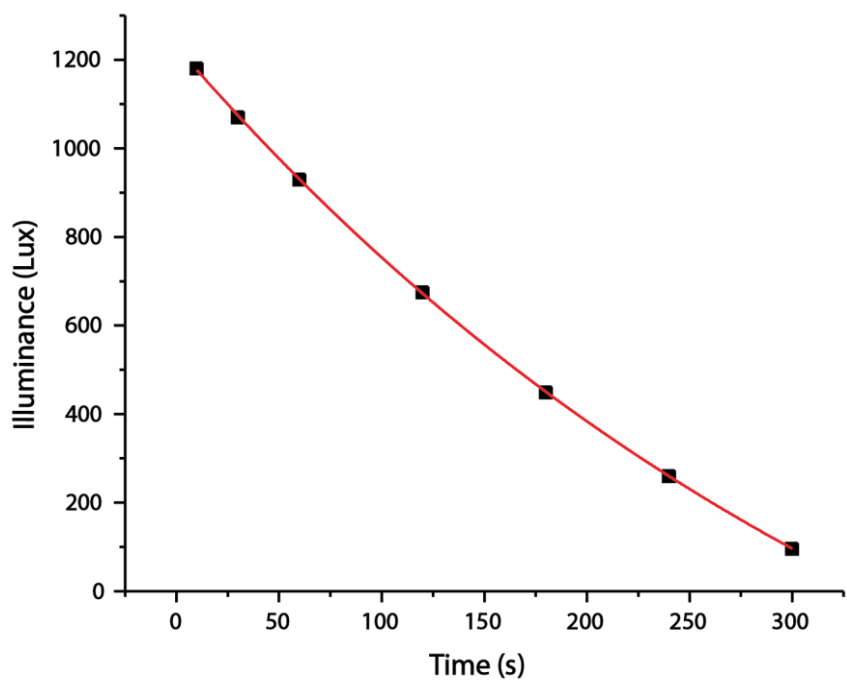

**Figure S4. Decay of brightness of green LED powered by paper-gel cells.** The green LED was connected to a stack of 16 paper-gel cells in series, identical to the stack depicted in Figure 3. The decay was fit with a single exponential decay (red line), yielding a half-life of 135 s.

**Supplementary Information 5. Rehydration and assembly of dehydrated paper-gels**

To use the paper-gel artificial electric organ as an on-demand power source, it would be beneficial to decouple the fabrication of paper-gels from their assembly and connection across a load. As stated in Supplementary Information 2, hydrogel evaporation lowers potential power output of a stack of gels by increasing electrical resistance. A strategy of taking hydrogels that have partially or completely evaporated and rehydrating them prior to assembly would, then, provide a method for quickly generating a functional power source when needed without keeping hydrogel precursor solutions at hand.

To test the feasibility of this type of rehydration strategy, we fabricated the paper-gels required for the assembly of two identical paper-gel cells with a salinity gradient of 1.5 M to 0.015 M CsCl. We chose CsCl instead of LiCl because a gel with a high concentration of hygroscopic LiCl would not dry as quickly or completely. We assembled one paper-gel cell immediately after completing the UV curing process and measured the maximum power output of the cell once every 60 s. We placed the other paper-gels in a laminar flow hood overnight to allow them to dry. The next day, we added an amount of water equivalent to the water content of each gel (in its freshly prepared state) to the surface of the gel, and waited for 15 min for the water to be absorbed. We then repeated the assembly and recording process of the first cell. The rehydrated cell achieved 55% of the maximum power output at time = 60 s (Fig. S5). Furthermore, the power output of the rehydrated cell remained stable for five minutes, which is atypical of freshly-made paper-gel cells.

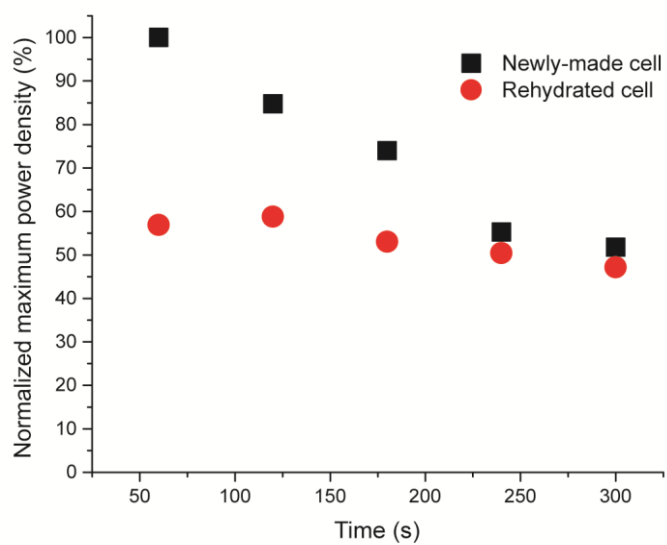

**Figure S5. Power output of rehydrated paper-gel cell.** The maximum power density of a paper-gel cell (black squares) assembled immediately after paper-gel fabrication was compared with the maximum power density of a paper-gel cell assembled from paper-gels that were dried overnight and rehydrated with water prior to assembly (red circles). The data were normalized by dividing the maximum power density at each time point by the maximum power density of the newly-made cell at time = 60 s. The rehydrated cell achieved 55% of the power output of a newly-made cell at time = 60 s. The power output of the rehydrated cell remained relatively stable over five minutes compared to the output of the newly-made cell.

**Supplementary Information 6. Calculating the maximum power output of a paper-gel stack**

The power $P$ (W) dissipated across a load resistor $R_L$ (Ω) is given by Equation S2,[6]

$$P = \frac{V_L^2}{R_L} \quad \text{(S2)}$$

where $V_L$ (V) is the voltage across the load. To calculate the maximum power output of a given stack of paper-gels, we first measured the open-circuit voltage of the stack. We then connected the stack in series with a load resistor of a known value. The resulting circuit was a simple voltage divider, with part of the generated voltage dropping across the load resistor and the rest dropping across the internal resistance of the stack $R_{int}$ (Ω). The amount of voltage across the load resistor was proportional to the value of the resistor over the total resistance of the stack, as given by equation S3.[6]

$$V_L = V_{OC} \frac{R_L}{R_L + R_{int}} \quad \text{(S3)}$$

From this equation, we can calculate the internal resistance of the stack. The maximum power transfer theorem states that maximum power is extracted from a DC power source when $R_L = R_{int}$. For a power source with a linear current-voltage relationship, this condition is met when $V_L = V_{OC}/2$, transforming Equation S2 into Equation S1, which we used to calculate maximum power. To confirm that the artificial electric organ displays a linear current-voltage (IV) relationship, we measured the voltage and current of a paper-gel cell connected to several different load resistances (Fig. S6).

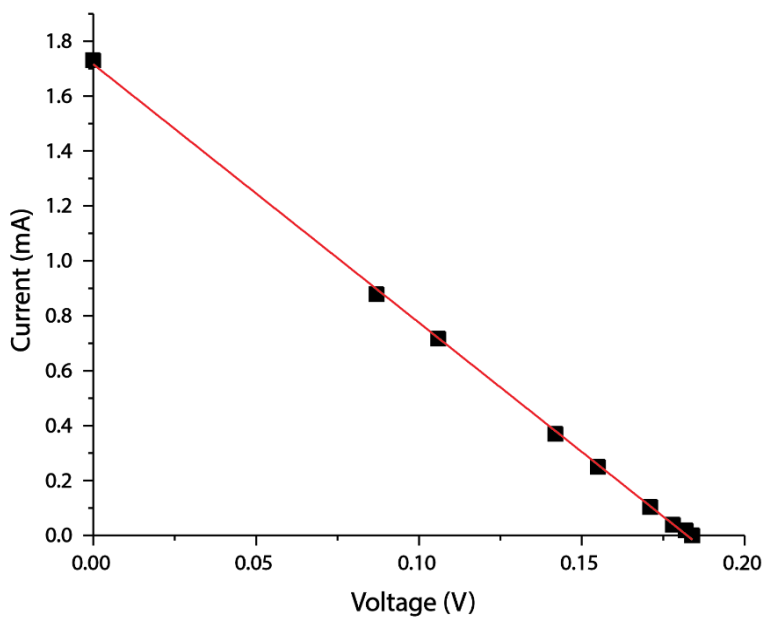

**Figure S6. Representative current-voltage (IV) relationship of paper-gel cell.** The voltage of a single paper-gel cell was measured as resistors of various magnitudes were connected in series. The x-intercept represents the open-circuit voltage, and the y-intercept represents the short-circuit current of the stack. The data were fit linearly (red line); the maximum power density was 0.31 W m$^{-2}$.